\documentclass[twocolumn,11pt]{article}
\usepackage[margin=2.5cm]{geometry}
\usepackage{url}
\usepackage{natbib}
\usepackage{graphicx}
\newcommand{\MODf}{\texttt{MOD-Finder}}

\newcounter{lastnote}

\title{MOD-Finder: Identify multi-omics data sets related to defined chemical exposure}

\author
{Sebastian Canzler,$^{1}$ J\"org Hackerm\"uller,$^{1,2}$ Jana Schor$^{1,\ast}$\\
\\
\footnotesize{$^{1}$Young Investigators Group Bioinformatics and Transcriptomics,}\\
\footnotesize{Department of Molecular Systems Biology, Helmholtz-Centre for Environmental Research - UFZ,}\\
\footnotesize{Permoser Stra{\ss}e 15, Leipzig, 04317, Germany}\\
\footnotesize{$^{2}$Bioinformatics Department, Universit\"at Leipzig, H\"artelstra{\ss}e 16-18,}\\
\footnotesize{04107 Leipzig, Germany.}\\
\\
\footnotesize{$^\ast$To whom correspondence should be addressed; E-mail: jana.schor@ufz.de.}
}

\date{}

\begin{document}

\twocolumn[
  \begin{@twocolumnfalse}
    \maketitle
    \begin{abstract}
      
      \textbf{Summary:} Integration of multi-omics
	data on chemical exposure of cells or organisms promises a more
	complete representation of the responding pathways than single omics
	data. Data of different omics layers, like transcriptome or proteome
	is deposited in different repositories. Additionally, precisely
	specifying a chemical of interest that was used in the exposure
	experiments suffers from different nomenclatures and non-uniquely
	mapping of chemical identifiers. The manual search for corresponding
	omics data sets of different layers for exposure with a chemical of
	interest is thus a tedious task. We have developed
	\MODf\ (Multi-Omics Data set Finder) to efficiently search for chemical-related omics data sets
	in several publicly available databases in an automated manner. A
	plain and simple presentation of the returned omics data sets is
	augmented with effect information that are assumed to be triggered
	by the chemical of interest.\\ \textbf{Availability and
	  Implementation:} \MODf\ is implemented in \texttt{R} using the
	\texttt{Shiny} package. The web service is available at
	\url{https://webapp.ufz.de/mod\_finder} and the source code under the GNU
	GPL v3 license at \url{https://github.com/yigbt/MOD-Finder.}\\
	\textbf{Supplementary information:} Supplementary data are available at
	\url{https://www.ufz.de/index.php?en=44919}
        \vspace*{20pt}
    \end{abstract}

\end{@twocolumnfalse}
]

\section{Introduction}

Exposure of cells or organisms to hazardous chemicals
triggers a series of molecular events that may eventually lead to an
adverse outcome. Understanding the mechanisms that propagate the
initial effect at the molecular level to subsequent molecular and
cellular effects and finally to the adverse outcome, i.e. defining
adverse outcome pathways, is a major field of research in
toxicology. Omics techniques have been instrumental for these analyses
and as cellular pathways combine biomolecules of different types,
integration of multi-omics data is increasingly considered in
toxicology, see e.g. \cite{Buesen:2017}.

Omics data are deposited in repositories that are specific for one
omics layer, e.g. transcriptomics, or proteomics and several
repositories may be used by the community for one layer. Assembling
corresponding multi-omics data thus is a tedious task that highly
depends on the quality of meta data. The \texttt{Omics Discovery
  Index} (OmicsDI) framework facilitates the aggregation of multiple
omics data sets across a variety of sources
\citep{Perez-Riverol:2017}.

Identifying corresponding multi-omics data sets for exposure with a
defined chemical of interest (CoI) remains a challenge also when using
\texttt{OmicsDI}: Chemicals exhibit a considerable degree of naming
variability, which may include different, in some cases non-unique
trivial and trademark names, acronyms, or identifiers following
standard nomenclature guidelines as provided by the IUPAC
\citep{Krallinger:2015}. The reliable \emph{digital} identification of a
CoI is therefore a prerequisite for assembling multi-omics
data. Annotated effect data of a chemical can assist in selecting the
right digital identifier for a CoI in case of naming or identifier
mapping collisions. To the best of our knowledge, a tool for
assembling multi-omics data for exposure with a CoI is currently not
available.

Here, we present \MODf\ (Multi-Omics Data set Finder). An easy-to-use
web application, to search for multi-omics data sets related to a user
defined CoI. It supports the reliable identification
of the correct chemical, provides integrated meta-information and
functional annotation of the queried chemical, and lists detected
multi-omics data sets from various sources for transcriptomic,
proteomic, and metabolomic data. \MODf\ visualizes described molecular
effects on gene and pathway level as well as related diseases.

\MODf\ provides known functional annotation of chemicals of interest
together with multi-omics data sets intended to be used in multi-omics
data analysis approaches in toxicology. Examples are benchmarking
multi-omics data integration software, evaluation of the benefit relative
to single-omics data analysis, hypotheses generation, and elucidation
of potential mode of actions.

\section{\MODf\ - Methods and workflow}

\begin{figure}
  \includegraphics[width=\columnwidth]{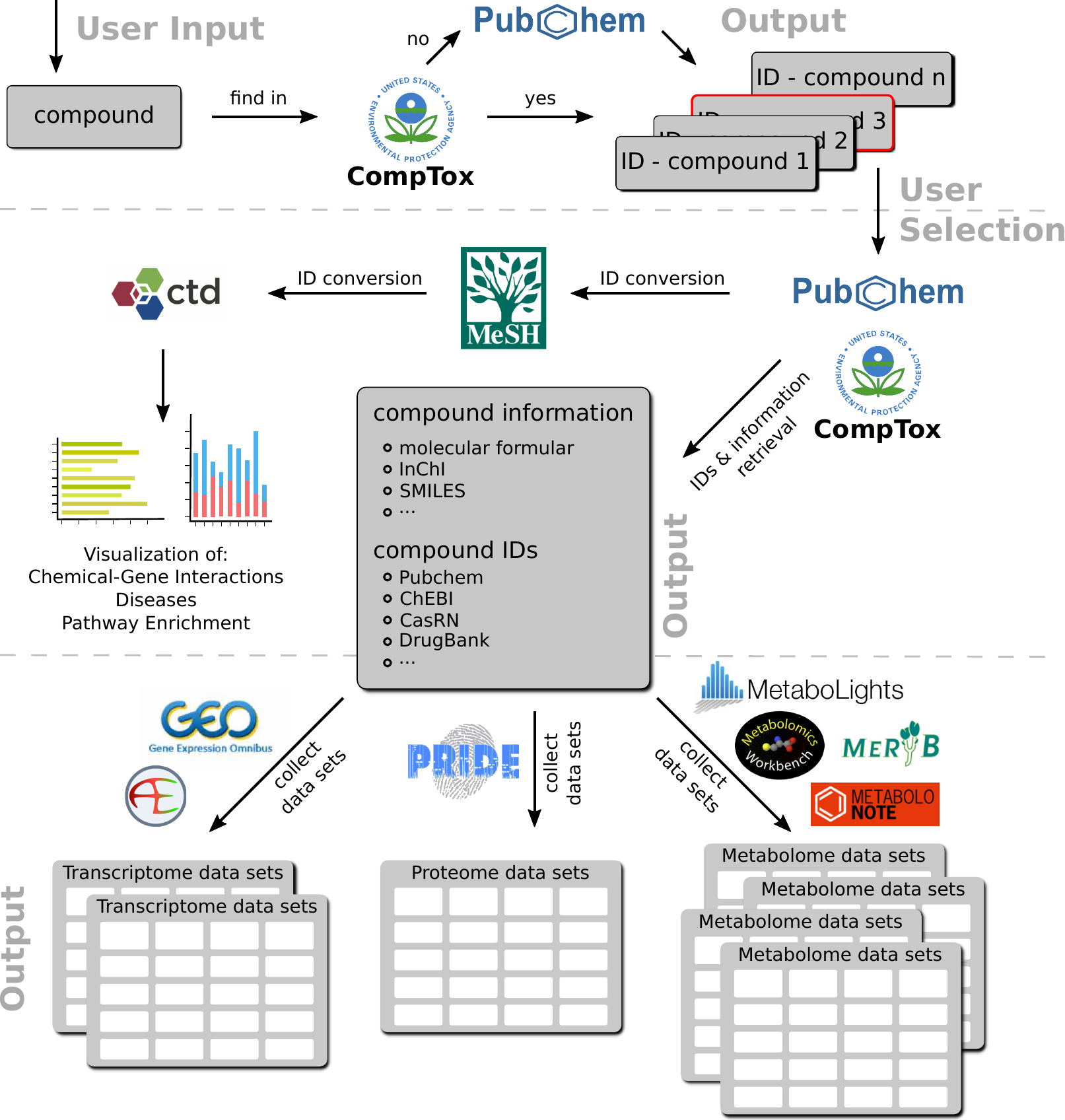}
  \caption{The \MODf\ workflow: top) \emph{digital} identification and
    selection of a CoI, middle) collection, integration, and
    visualization of chemical related information and effect
    annotation, bottom) search of related omics data sets.}
  \label{fig:workflow}
\end{figure}

The workflow is split into three parts that are visualized in Figure~\ref{fig:workflow}.

\paragraph{\emph{Digital} identification of the chemicals of interest.}
CoIs are referenced using several independent digital identifiers,
which often cannot be mapped to one another straightforwardly.
Many-to-many and many-to-none relations are common. The first step in
\MODf\ considers the correct identification of the CoI provided by the
user. We implemented a search strategy that queries all provided
identifiers, or text-based entries in the \texttt{CompTox Chemistry
  Dashboard}\ \citep{Williams:17}, a source of high-quality,
structure-curated, open chemical data, for a number of matching
CoIs. For CoIs not found in \texttt{CompTox}, \MODf\ applies a similar
search strategy in the
\texttt{Pubchem}\footnote{https://pubchem.ncbi.nlm.nih.gov/}.

\paragraph{Functional annotation of chemicals of interest.}
Prior knowledge on biological effects of the CoI may facilitate
selection of relevant data sets. \MODf\ therefore queries the
Comparative Toxicogenomics Database
(\texttt{CTDbase})\ \citep{Davis:19} for known relations between the
CoI and genes, pathways, and diseases and visualizes this data.

\paragraph{Query omics data bases.} 
\MODf\ queries \texttt{NCBI\ GEO} \citep{Barrett:2013} and 
\texttt{ArrayExpress} \citep{Kolesnikov:2015}, 
for gene expression data sets (transcriptome). 
The \texttt{PRIDE} database
\citep{Vizcaino:2016} is queried as one of the most prominent data
repositories of mass spectrometry based proteomics data and the
\texttt{MetabolomeXchange} \footnote{http://www.metabolomexchange.org}
international data aggregation and notification service for
metabolomics. Different data base layouts, degrees, and types of access
to data, e.g. RESTful APIs or downloadable SQL data base files, are
handled appropriately by \MODf.

\paragraph{Availability and maintenance.} 
\MODf\ is provided as an \texttt{R\ Shiny} application hosted at
\url{https://webapp.ufz.de/mod_finder}. The source code is available
at \url{https://github.com/yigbt/MOD-Finder.}  \MODf\ is updated
periodically and upgraded to include further public omics databases or
additional omics layers.  \MODf\ releases are tracked and separately
available at \texttt{github.}

\paragraph{}
\MODf\ presents a comprehensive, clearly structured overview
of the identified data sets using visuals and customizable
tables. Links to subsequently download the data sets are provided
within the app in separate output panels.

\section{Conclusion and discussion}
\label{discussion}

Unraveling the pathway response to chemical exposure is a
major aim of current toxicological research. Integration of
multi-omics data promises a much more complete representation of
pathways and therefore supports not only a deeper mechanistic
understanding of toxicological processes, but also facilitates
predictive approaches to chemical risk assessment \citep{Buesen:2017}.
\MODf\ simplifies the entry into such data-intensive approaches by
enhancing the systematic omics data set retrieval and providing prior
knowledge of known effects of the chemicals in a clear visualization.
This combination is unique and makes \MODf\ a useful
tool for multi-omics data approaches in toxicology. To the best of our
knowledge, the only other publicly available multi-omics data
retrieval tool is \texttt{OmicsDI}. It integrates information on
different biological entities, including genes, proteins, metabolites,
and the corresponding publications from PubMed but lacks integration
of chemical knowledge.  \MODf\ has a clear focus on retrieving
multi-omics data for chemical exposure experiments. This includes the
correct computational identification of a chemical of interest,
knowledge retrieval of described effects combined with the selection
of related omics data sets from transcriptomics, proteomics, and
metabolomics measurements.

\section*{Funding}
This work was funded in part by the Cefic Long-Range Research
Initiative Program (Project C5-XomeTox).

\bibliographystyle{natbib}

\bibliography{MODfinder_main}

\end{document}